\documentclass[12pt,preprint2]{aastex}

\def\Term#1 #2 #3/{\mbox{$\,^{#1}\!#2_{#3}$ }}
\def\Termo#1 #2 #3/{\mbox{$\,^{#1}\!#2^o_{#3}$ }}
\def\sterm #1 #2 #3/{\mbox{$\,_{#3}\!^{#1}\!#2$}}

\begin{document}

\title{Breit-Pauli energy levels, transition probabilities, and lifetimes for $3d^5$ levels in 
\ion{Fe}{4} of astrophysical interest}

\author {Charlotte Froese Fischer\altaffilmark{1}}
\author{Robert H. Rubin\altaffilmark{2,3}}

\email{Charlotte.F.Fischer@Vanderbilt.Edu; rubin@cygnus.arc.nasa.gov}

\altaffiltext{1} {Department of Computer Science, Box 1679B \\ Vanderbilt University, Nashville
TN 37235, USA}
\altaffiltext{2} {NASA/Ames Research Center, Moffett Field, CA
94035-1000, USA}
\altaffiltext{3} {Orion Enterprises, M.S. 245-6, Moffett Field, CA
94035-1000, USA}

\def\Termo#1 #2 #3/{\mbox{$\,^{#1}\!#2^o_{#3}$ }}

\date{\today}

\begin{abstract}
Energy levels, lifetimes, and transition probabilities for 
transitions between computed levels of $3d^5$ of \ion{Fe}{4} are reported.
The E2 and M1 transition probabilities are compared with earlier theoretical results, 
often only the values published by Garstang in 1958.
 From the available astronomical observations of optical emission lines arising from the
same level, a few direct tests are now possible and they show consistency with the 
theoretical calculations.

\end{abstract}


\vspace{12pt}

\section{Introduction}

	Triply ionized iron Fe$^{3+}$ is expected to be a significant
fraction of gaseous iron in many nebulae.
Indeed, Fe$^{3+}$ may be the dominant ionic state of Fe
in many nebulae, including 
\ion{H}{2} regions and planetary nebulae (PNs).
For instance, in two independent photoionization models
of the benchmark Orion Nebula, the fractional ionization
$<$Fe$^{3+}$$>$ was predicted to be 0.744 by Baldwin et~al.\ (1991)
and 0.533 by Rubin et~al.\ (1991a, b).
Naturally, in order to get a grip on iron abundances,
it is important to 
treat the dominant ionic component.
To interpret observations of emission lines
of [\ion{Fe}{4}], it is necessary to have reliable atomic data,
including effective collision strengths  (Fe$^{3+}$ with electrons)
and transition probabilities (Einstein $A$-values).
These atomic data are important for interpreting astronomical
observations over a wide spectral range from the ultraviolet to the infrared.

	In our earlier paper
(Froese Fischer \& Rubin 1998),
we provided $A$-values for transitions between the 12-lowest 
energy levels.
Effective collision strengths had just become available for these
12-lowest levels
(Berrington \& Pelan 1995, 1996).
Although there were $A$-values available from 
Garstang (1958),
improvements in the state-of-the-art made it worthwhile to
recalculate a complementary set that 
would permit a solution for the detailed population statistical
equilibrium for the 12-level atom. 
This depends on the electron density ($N_e$) and electron temperature ($T_e$).

One of the motivating factors in our 1998 paper was to provide improved
$A$-values for a set of transitions involved in
the determination of the intensity of 
the UV [\ion{Fe}{4}]  2836.56 \AA\
line measured with 
the Hubble Space Telescope
(HST) in the Orion Nebula by Rubin et~al.\ (1997)~--
the first detection of an [\ion{Fe}{4}] line in an \ion{H}{2} region.
They had measured the flux of 
[\ion{Fe}{4}] 
($3d^5~^4P_{5/2}$ 
$\rightarrow$ 
$3d^5~^6S_{5/2}$)
$\lambda_{vac}$~= 2836.56~\AA\ 
and set an upper limit on the sum of fluxes
of [\ion{Fe}{4}] 
($3d^5~^4D_{5/2,3/2}$ $\rightarrow$ $3d^5~^6S_{5/2}$) 
$\lambda_{vac}$~= 2568.4, 2568.2 \AA.

	Unfortunately, Fe$^{3+}$ does not have intrinsically bright lines
under nebular conditions. 
Recently there have been measurements of some optical lines 
of [\ion{Fe}{4}]
(e.g., Rodr\'\i guez 2003).
Rodr\'\i guez found for five 
nebulae
that the Fe$^{3+}$ abundance derived from 
observations of 
[\ion{Fe}{4}]
lines was systematically lower than expected.  
This is in the same direction as found earlier
by Rubin et~al.\ (1997) for the Orion Nebula.

In their study of the bipolar PN
Mz~3, Zhang \& Liu (2002)
measured 
five optical [\ion{Fe}{4}]  lines as well as 
several lines of [\ion{Fe}{3}].  
They found evidence for 
high $N_e$ in the Fe$^{++}$ region
of log~$N_e$(cm$^{-3}$)~= 6.5. 
They also suggested that $N_e$ in the central emitting core
could be even higher.  
An interpretation of those data would benefit from improved $A$-values,
particularly when dealing with higher density gas where 
the statistical equilibrium for the energy level populations
depends critically on the transition rates.

	Most, if not all, of the optical 
[\ion{Fe}{4}] lines observed astronomically arise
from energy levels above the 12-lowest levels for which we calculated
$A$-values (Froese Fischer \& Rubin 1998).   
These optical, as well as IR, lines originate from energy 
levels above the $^4D$-levels (beginning with level 13).

Zhang \& Pradhan (1997)
extended the calculation for
[\ion{Fe}{4}] effective collision strengths to a 140-level atom (49 terms
up to 15 Ryd).
Because of the availability of these data and the need 
to consider
higher levels to interpret/predict most of the [\ion{Fe}{4}]  lines
that are being observed, we are making these new, improved
$A$-value computations including 16 terms, which comprise the 
37 lowest-lying energy levels (below 1 Ryd).

The goal of our earlier publication was to predict transitions probabilities in emission from $3d^5$ \Term 4 G {}/,
\Term 4 P {}/, and \Term 4 D {}/ to the \Term 6 S {5/2}/ ground state.  
In the process, transition probabilities were also computed for transitions between the levels of 
the \Term 4 L {}/ terms.  
Four theoretical methods were compared and best estimates identified.  
Frequently these were the multiconfiguration Hartree-Fock with Breit-Pauli (MCHF+BP) results, though not always.  
The multiconfiguration Dirac-Hartree-Fock with the Breit (MCDHF+B) correction was selected in some cases for 
magnetic dipole (M1) transitions. 
In a few instances,  a semi-empirical method based on orthogonal operator 
techniques 
(Raassen \& Uylings 1996)
was selected. 
In this paper we extend the Breit-Pauli work to include all levels of the 
$3d^5$ configuration 
up to 1 Rydberg.

\section{Computational Procedure}

In the Breit-Pauli approximation, the wave function, $\Psi$, is a linear
combination of configuration state functions (CSF) 
of the  form
\begin{equation}
\Psi ( \gamma J ) = \sum_{LS} \sum_j c_j(LSJ) \Phi (\gamma _j L S J ),
\end{equation}
where $\gamma$ usually represents the dominant configuration and any additional
quantum numbers required for uniquely specifying the state.
The CSFs, $\Phi(\gamma_j LSJ)$, for a configuration and coupling
$\gamma_j$,  term $LS$, 
and total angular momenta $L$ and $S$ coupled to $J$, are built
from a basis of one-electron spin-orbitals,
\begin{equation}
\phi _{nlm_lm_s} = \frac{1}{r} P_{nl} (r) Y_{lm_l} (\theta, \varphi)
\chi_{m_s}.
\end{equation}
The expansion coefficients, $c_j(LSJ)$, and the corresponding energy,
$E(LSJ)$, are an eigenvector and eigenvalue, respectively,
of the interaction matrix of these CSFs
as defined by the Breit-Pauli Hamiltonian.  The evaluation of the matrix elements 
is considerably simplified if the spin-orbitals are orthonormal, which
means that, for a given $l$, the radial functions $P_{nl} (r)$ must be
orthonormal, and that the same radial functions must be used to represent the
different terms.

In this approximation, the configuration states in the expansion and the spin-orbitals
are determined from non-relativistic calculations.
The multiconfiguration Hartree-Fock (MCHF) method was
used for this purpose, extended to obtain the ``best'' radial functions for 
the set of $LS$ terms.  In all the present
work $1s^22s^22p^6$ was considered an inactive core.  

Typically, the different terms of a configuration that interact through relativistic operators
for one or more $J$ values are grouped and the radial functions optimized for the group. 
But many of the transitions among the levels of $3d^5$ are spin-forbidden.  
These transitions arise from small admixtures of quartet components to doublet levels and vice versa. 
Thus even small admixtures required terms to be in the same group. 
What compounded the difficulty in this case was the fact that
\sterm 2 D 5/ CSF interacts strongly through a coulomb interaction with the \sterm 2 D 1/ CSF. (In this
notation, the subscript preceding the $LS$ term is the seniority of the term.  We will include seniority only
when it is necessary to distinguish different terms.)  
The former is the dominant component of a term low in the spectrum whereas the latter is a dominant 
component for a term high in the spectrum.  
At the same time, the \sterm 4 F {}/ and \sterm 2 F 3/ interacted strongly, as well as others.  
In order not to miss small relativistic interactions, it was decided to include all levels in one group, 
but omit optimization of orbitals for the ground state which contains considerably less correlation than 
the other levels.

With this in mind, a multi-reference set was created for each term that included the $3p \rightarrow 4p$ 
and $3d \rightarrow 4d$ replacements from $3s^23p^63d^5$ to allow for term dependence of the $3p$ and $3d$ orbitals. 
Calculations were performed with orbital sets of increasing size with the $n=4$ orbital set including all 
occupied orbits as well as $4s,4p,4d,4f$.  
The $n=4$ expansion for a given term was obtained through single (S) and double (D) excitations from 
the multireference set to the $n=4$ orbital set.   For $n=5,6,7$ the $3s$ shell was kept
inactive, and SD excitations with at most one excitation from the $3p^6$ subshell were added to the $n=4$ 
expansion.  
Given that each of these expansions was fairly long and that there are 13 terms, 
it was necessary to eliminate the small contributors to an $LS$ expansion. 
MCHF calculations were performed for each $LS$ term and expansions condensed in that configuration state 
functions with an expansion coefficient less than 0.00001 were eliminated.
Once these expansions had been obtained, simultaneous optimization was performed for all twelve terms 
(omitting \Term 6 S {}/) to obtain radial functions.

Once the radial basis had been determined, the Breit-Pauli interaction matrix was determined 
including all $LS$ terms. The results define our {\it ab initio} energies.  

When observed energy level data 
are
available, transition probabilities, 
$A_{ki}$, can be improved by various energy adjustments.
The first such adjustment is one that corrects for the transition energy.
Let $r = {\Delta E_{obs}}/{\Delta E_{calc}}$.  Then $A_{ki}(adj) = r^m
A_{ki}(calc)$, where $m=3$ for M1 transitions and $m=5$ for 
electric quadrupole (E2) 
transitions.  This adjustment is the most straight forward, but for
Breit-Pauli calculations where we have the mixing of different $LS$ terms,
this mixing itself is affected by the ``term energy separation.''  Though it is
possible, in simple cases, to correct a computed $A_{ki}$ for such an error,
it is simpler to first adjust the $LS$ term energies in a Breit-Pauli 
calculation so that, for selected $J$ values, the separation of terms
is in close agreement with observed. Typically, this is done by first 
determining energies without an adjustment, noting the deviation from 
observed for the $J$ values of different terms, selecting a $J$ value from each
term for adjusting all levels of a term, and then modifying
the diagonal energies of all CSFs associated with a given $LS$ by this
amount. Unless selected $J$ values are exceedingly close, 
just one such iteration brings energies into
close agreement with observed.  It usually does not change the ``spread'' of a 
term, the difference in energy between the highest and lowest level of a term.
But $3d^5$ being a half-filled shell, there is no diagonal spin-orbit interaction 
and fine-structure splitting is more complex.
We refer to calculations where only the diagonal energies have been changed
as ``adjusted".  
Because there are as many as three terms with the same $LS$ value, as for the 
\Term 2 D {}/  (\sterm 2 D 5/, \sterm 2 D 3/, 
and \sterm 2 D 1/, where the 
preceding subscript is the seniority), the adjustments were done in groups in order of energy: 
\begin{enumerate}
\item \Term 6 S {}/  -- \Term 4 F {}/
\item \sterm 2 H 3/ -- \sterm 2 F 5/
\item  \sterm 2 S {}/ -- \sterm 2 D 3/
\item  \sterm 2 G 3/ -- \sterm 2 D 1/
\end{enumerate}

Table 1 reports the final spectrum, the splitting relative to the lowest level of the 
$LS$ term, and the
difference with observed
(NIST Atomic Spectra Database, Physical Reference Data, http://physics.nist.gov). 
Since the adjustment involves only one $J$ of a term, not all levels are in close agreement with observed. 
The largest deviation from the observed spectrum was 115 cm$^{-1}$ in an excitation energy of 50,051 cm$^{-1}$.

\section{Analysis of energy levels}

Though the differences in our energies relative to 
observed
are not large, 
they do indicate that the fine-structure splitting has not been determined accurately.
If a term is simply shifted relative to the ground state, then the difference with observed should be 
essentially constant.
In the quartet terms, the levels are not always in the correct order. 
For example, the observed order of levels of \Term 4 G {}/ is (11/2, 9/2, 5/2, 7/2), 
the present order is (11/2, 5/2, 7/2, 9/2) whereas the order in Garstang's 
semi-empirical work was (5/2, 7/2, 11/2, 9/2). 
All levels are close together, with the observed spread being only 
$\sim$60 cm$^{-1}$.  
However, the splitting within a multiplet is only important for the transitions within a multiplet, 
determining the order and the wavelength.  
The length form of the line-strength (S)
is largely independent of the energy.

Table~1 shows that the wave functions for many levels have a composition that includes a number of 
$LS$ terms
at the 1\% level.  The first number is the percentage composition of the dominant configuration state
function.  Many are listed as 95-96\%.  The remaining composition represents correlation effects and small
relativistic interactions.  But some levels have a highly mixed 
composition 
and the accuracy of transition
probabilities, particularly spin-forbidden transitions, from such levels depends on how well the composition
of the wave function is represented.
The accuracy of this mixing  can be assessed to some extent by the accuracy of the 
spectrum for a particular $J$ and the separation 
between 
levels.  An important value is $J=5/2$ and in 
Table~2 these energy levels are listed along with the separation of a 
particular level from the previous one.  The present adjusted energy separations
are compared with observed separations and those derived by 
Garstang (1958). 
Considering the incomplete identification of the spectrum at the time of 
Garstang's work, his results are remarkable.  
Table~2 shows immediately that a potentially 
strong mixing of the \sterm 2 F 3/ and 
\Term 4 F {}/ configuration states may occur for $J=5/2$, with a separation of only 
671 cm$^{-1}$ between the energies of these two terms. 
This separation has been reproduced with an error 1.5\%,  though it needs to be remembered
that these are adjusted values.  The {\sl ab initio} separation was 894 cm$^{-1}$. 

Close levels of the same $J$ are sensitive to the separation and so this can be used as a test for
accuracy, but strong mixing can also occur simply because of a large off-diagonal matrix element 
in the Breit-Pauli Hamiltonian. In Table~1, we note the strong mixing of the \Term 4 F {}/ 
with the nearby \sterm 2 F 3/ term for $J=5/2$, but the strongest mixing is between \sterm 2 D 5/ and  
\sterm 2 F 3/ 
for $J=5/2$ though this separation is larger.

\section{Comparison of calculated transition probabilities}

The complete set of transition probabilities between all the levels of
our energy adjusted data can be found at 
{\tt http://atoms.vuse.vanderbilt.edu}.\break
Transition probabilities between many levels are reported in the
Appendix with the wavelengths 
corrected to agree with observed and the 
oscillator strengths ($f_{ik}$)
and $A_{ki}$ 
modified accordingly. 
This latter correction was only important for transitions within a multiplet.
The Appendix is available as a machine-readable Table in the electronic 
edition.

In order to assess the accuracy of these results, we compare the present transition probabilities between
the \Term 6 S {}/, \Term 4 P {}/, and \Term 4 D {}/ for which our previous publication had compared four 
different theoretical approaches:  Garstang's early calculations
(Garstang 1958),
the semi-empirical
orthogonal operator method
(Raassen \& Uylings 1996),
an MCHF with Breit-Pauli (MCHF+BP) method,
and a multiconfiguration
Dirac-Hartree-Fock method with a Breit (MCDHF+B) correction, and the most reliable value 
identified.  E2 transition probabilities are compared in 
Table~3
and M1 transition probabilities in 
Table~4.
Though our previous calculations were not as ambitious as the present, MCHF+BP was often identified as the
most reliable, but for some M1 
transitions 
MCDHF+B was selected and for some E2, the results by 
Raassen \& Uylings (1996).
The present results are often similar to the 
earlier 
ones, but some small values 
are now closer to those of 
Raassen \& Uylings, possibly because more term mixing was taken into account.
An example is the \Term 6 S {5/2}/ -- \Term 4 G {7/2}/ E2 
value which previously was computed to have a
transition probability of 5.27$\times 10^{-12}$ has now become 
3.32$\times 10^{-8}$. The latter is in excellent agreement
with the orthogonal operator value of 3.18$\times 10^{-8}$ that fortuitously had been selected as the 
most reliable.  

	For the transition probabilities for the newly included terms, the 
only other values are those reported by Garstang (1958)
in his Table~III, but only up to level 33.  
In Table~5 we present only those multiplets arising from higher levels than
those shown in Tables~3 and 4 and that give rise to an optical emission line 
that may possibly have been seen astronomically.
As shown in 
Table~5, for spin-allowed transitions, some results are in excellent agreement
with the Garstang values.
	There are five transitions for which very small
$A_{ki}$ values 
(no larger than 3.41$\times 10^{-14}$~s$^{-1}$)
are given in Garstang but not in the Appendix.
We calculate that all of these are now less than
5.3$\times 10^{-14}$~s$^{-1}$ and negligible.
For these or other
transitions not listed in the Appendix,
they are available from
{\tt http://atoms.vuse.vanderbilt.edu}.\break
Again, the web site values have not been adjusted to observed
wavelengths, and the steps mentioned in section~2
should be followed to modify $A_{ki}$. 

In conclusion, by including the Breit-Pauli interactions between all
the terms of $3d^5$, the calculations are more complete.  
Some E2 transition probabilities for transitions
among the four lowest terms remain similar but others have come closer
to semi-empirical values. An example is the $^4\!G - ^4\!P$ multiplet. 

\section{Comparison with observations}

	Because of the inherent faintness of the
[\ion{Fe}{4}]  lines, there is as yet very little data for a
``direct" test of the $A$-values.  
One astronomical object that permits such a test is 
RR~Telescopii, which is a symbiotic nova
(e.g., McKenna et~al.\ 1997).
In Table~1 of their paper, 
they have observations of two emission lines arising on the same level,
for each of two different upper levels.
All are within the $^4F \rightarrow$ $^4G$ multiplet.
The first test is the intensity ratio of the 
4903.07~\AA\ line
to the 
4899.97~\AA\ line (rest air wavelengths from NIST).
(These are listed somewhat differently in the McKenna et~al\ table as 
4903.50 and 4900.05.)
These are respectively 
$^4F_{7/2} \rightarrow$ $^4G_{7/2,~9/2}$.
To facilitate comparison with the theoretical transition probabilities
in our Tables, the vacuum wavelengths for these lines are 4904.44
and 4901.34, respectively.

	The predicted line intensity ratio\break
I(4903)/I(4900) is simply
the ratio of the products of the line frequency and the $A$-value,
where the $A$-value is the sum of E2 and M1.
According to the Appendix, the present predicted ratio
is 1.10 which is close to that predicted by Garstang (1958)
of 1.04.   The observed ratio from McKenna et~al. (1997) is 1.13,
where F(4903)~= 4.4 and F(4900)~= 3.9 in units of
10$^{-14}$~ergs~cm$^{-2}$~s$^{-1}$.

	The second pair of observed lines in RR~Tel 
permits a test of the intensity ratio of I(4907)/I(4918).
The best (NIST), rest air wavelengths are 4906.56 and 4917.98~\AA\
(these are listed somewhat differently in the McKenna et~al.\ table as 
4906.70 and 4918.10)
and are respectively 
$^4F_{9/2} \rightarrow$ $^4G_{11/2,~9/2}$
with vacuum wavelengths of 4907.93 and 4919.35.
	The present predicted line intensity ratio
I(4907)/I(4918) from the Appendix is 3.17,
while the ratio from Garstang (1958) is 2.72.
The observed ratio from McKenna et~al. (1997) is 2.39,
where F(4907)~= 9.1 and F(4918)~= 3.8~$\times$
10$^{-14}$~ergs~cm$^{-2}$~s$^{-1}$.
However, this comparison does not account for a possible contribution of
an \ion{Fe}{3} line at 4918.00~\AA\ listed also as a possible identification
along with the 
[\ion{Fe}{4}] line for their observed feature at 4918.04~\AA\
(McKenna et~al.\ 1997). 
That would lower both of the above predicted ratios some
undetermined amount.

	For one of the PNs that Liu et~al.\ (2004)
observed, NGC~6884, two emission lines of 
[\ion{Fe}{4}] from the same upper level
are listed in their Table~3.
These are the same pair of lines as was discussed above
in the first instance for RR~Tel.
However, for NGC~6884, they list also as a possible identification
for their observed feature at 4903.32~\AA, a line due
to [\ion{Fe}{6}] at 4903.30.
This presumably is the $^2F_{5/2} \rightarrow$ $^2P_{3/2}$ transition,
which would be blended with any putative [\ion{Fe}{4}] 4903.07.

	We predict I(4903)/I(4900)~= 1.10
while Garstang transition probabilities predict 
nearly the same ratio (1.04).
Liu et~al.\ have observed fluxes
F(4899.21)~= 0.012 and F(4903.32)~= 0.039
scaled to F(H$\beta$)~= 100.
Hence from the observed flux ratio (3.2), one would conclude that the 
[\ion{Fe}{6}] 4903.3 is the major contributor to the observed blend
(subject to observational uncertainties).
There are indeed other [\ion{Fe}{6}] lines they list as observed in NGC~6884.
Besides the RR~Tel data, to the best of our knowledge, this 
is the only other instance of astronomically observed [\ion{Fe}{4}]  multiple 
lines that originate from the same level and thus in principle allow a direct test 
of the calculated $A_{ki}$ ratio.
However, in this case, the test is not a clean one because
of the likely blending of the [\ion{Fe}{4}]  
line with an [\ion{Fe}{6}]  line.

	It is beyond the scope of this paper to extract additional information
from 
the current rather sparse number 
of [\ion{Fe}{4}] lines observed in a given nebula. 
To do this involves a detailed population
statistical equilibrium set of equations that uses effective 
collision strengths as well as transition probabilities to
predict [\ion{Fe}{4}] line intensities as a function of $N_e$ and $T_e$.
By providing this improved set of $A$-values, the ingredients 
to do such a calculation are in place.
A machine-readable
file is available, by contacting either author by email,
that provides the sum of the M1~+~E2 transition probabilities for
the entries in the Appendix.

\begin{acknowledgments}
This work was supported by the Chemical Sciences, Geosciences and
Biosciences
Division, Office of Basic Energy Sciences, Office of Science, U.S.
Department of
Energy.
Support for RHR was from the NASA Long-Term Space Astrophysics (LTSA)
program.
We thank Dafna Bitton for assistance with preparing the Tables,
M\'onica Rodr\'\i guez for helpful comments, 
and Xiao-wei Liu for an advance  copy of Liu~et~al.\ (2004).

\end{acknowledgments}

\def\Term#1 #2 #3/{\mbox{$\,^{#1}\!#2_{#3}$ }}
\def\Termo#1 #2 #3/{\mbox{$\,^{#1}\!#2^o_{#3}$ }}
\def\sterm #1 #2 #3/{\mbox{$\,_{#3}\!^{#1}\!#2$}}

\clearpage


\columnsep=.2in
\begin{table*}
\caption{ Energy levels, their splitting, difference from
observed (computed - observed
(NIST)), lifetimes, and composition of the 
$3d^5$ levels.}
\label{spectrum}
\begin{center}
\begin{tabular} { r r r r r r r l}
\hline \hline
\multicolumn {1}{l}{$LS$}
&\multicolumn {1}{c}{$J$}
&\multicolumn {1}{c}{Level (cm$^{-1}$)}
&\multicolumn {1}{c}{Split.}
&\multicolumn {1}{c}{\quad Diff.\quad}
&\multicolumn {1}{c}{$\tau$ (s) }
&\multicolumn {1}{c}{\quad\quad}
&\multicolumn {1}{c}{Composition (\%)}\\

\hline 
 ${\!^6\!S}$ 
&                      5/2   &         0&             &  & && 96\\
${\!^4\!G}$ &  11/2  &     32245.54 &         & 0.04 & && 96\\
&                  5/2   &     32260.47 &       14.92 & -40.73 & 4.653e+04&& 96 \\ 
               &   7/2   &     32300.52 &       54.98 &  -5.18 &  3.492e+05 && 96 \\ 
               &   9/2   &     32307.06 &       61.52 &  14.26 &  1.216e+05 && 96\\

${\!^4\!P}$ 
&                  5/2   &     35229.91 &             & -23.87 & 6.408e$-$01 && 91 + 4\sterm 4 D {}/\\ 
               &   3/2   &     35362.69 &      132.78 &29.39 &  9.803e$-$01 && 92 + 3\sterm 4 D {}/\\ 
               &   1/2   &     35399.47 &      169.56 & -7.13 &   4.177e+05 && 95 + 1\sterm 4 D {}/\\

${\!^4\!D}$ 
&                   7/2   &     38778.87 &             & -0.53 &   1.955e+01 && 95\\ 
                &   1/2   &     38918.56 &      139.69 & 21.86 &   8.541e+00 && 95 + 1\sterm 4 P {}/\\ 
                &   5/2   &     38968.60 &      189.73 &  33.50 &  9.896e+00 & & 91 + 4\sterm 4 P {}/\\
                &   3/2   &     38973.28 &      194.41 &   2.68 & 9.395e+00  && 92 + 3\sterm 4 P {}/\\ 
  
${\!^2\!I}$ 
&                   11/2  &     47081.28 &             & 73.74 &   1.014e+02 && 95\\ 
                &   13/2  &     47164.24 &       82.96 &  42.34 &  3.280e+03 && 96 \\ 
  
$_5\!{\!^2\!D}$ &    5/2   &     49583.84 &             & 114.59 & 9.332e$-$01 && 52 + 25\sterm 2 F 3/ + 19\sterm 2 D 1/\\ 
                &    3/2   &     50165.99 &      582.15 &  1.25 &  2.145e+00 && 69 + 22\sterm 2 D 1/ + 5\sterm 4 F {}/\\

$_3\!{\!^2\!F}$  &   7/2   &     51395.45 &             &  65.84 &  1.211e+00 && 92 + 1\sterm 4 F {}/ +  1\sterm 2 F 5/\\ 
                &    5/2   &     52232.54 &      837.09 &  -9.06 &  1.085e+00 && 60 + 18\sterm 4 F {}/  + 13\sterm 2 D 5/ + 4\sterm 2 D 1/\\

 ${\!^4\!F}$ &   9/2   &     52611.64 &         &  -9.06 & 1.432e+00 && 94  + 2\sterm 2 G {}/\\ 
                 &   7/2   &     52712.37 &      100.74 & 16.97 &   1.464e+00 && 93 + 1\sterm 2 F 3/\\ 
                 &   5/2   &     52893.99 &      282.35 & 55.99 &  1.071e+00 && 76 + 10\sterm 2 F 3/ + 7\sterm 2 D 5/ + 2\sterm 2 D 1/\\ 
                 &   3/2   &     52894.38 &      282.75 & 57.28 &  1.219e+00 && 91 +  4\sterm 2 D 5/\\

 ${\!^2\!H}$  
                 &   9/2   &     56093.94 &             & 35.64 &   1.992e+00 && 79 +  16\sterm 2 G {}/\\ 
                 &  11/2  &     56421.86 &      327.91 & 53.06 & 9.092e$-$01 && 95\\

$_5\!{\!^2\!G}$ &    7/2   &     57411.28 &             &  3.28 &  1.138e+01 && 95 \\ 
                 &   9/2   &     57743.98 &      332.70 & 22.78 &   4.590e+00 && 78 + 17\sterm 2 H 3/ + 1\sterm 4 F {}/\\

$_5\!{\!^2\!F}$ &    5/2   &     61129.72 &             &  -26.78 &  3.790e+00& & 95\\ 
                 &   7/2   &     61258.00 &      128.28 &   3.60 & 3.502e+00 && 94 + 1\sterm 2 F 3/\\

${\!^2\!S}$ &    1/2   &     66720.76 &             &  0.66 & 1.024e+00 && 95\\

$_3\!{\!^2\!D}$ &    3/2   &     74098.16 &             & 1.56 & 6.529e$-$01 && 95\\ 
                 &   5/2   &     74147.52 &       49.36 & 14.42 & 5.050e$-$01 && 95\\

$_3\!{\!^2\!G}$  &   9/2   &     82895.92 &             & 1.02 &1.543e$-$01 && 95 \\ 
                 &   7/2   &     82943.02 &       47.10 & 45.72 & 1.458e$-$01 && 95\\

${\!^2\!P}$ &    3/2   &    100118.46 &             & 0.46 & 4.620e$-$02 && 95 \\ 
                 &   1/2   &    100119.61 &        1.14 & -6.34 & 4.512e$-$02 && 95 \\

$_1\!{\!^2\!D}$ &    5/2   &    108220.14 &             & -21.96 & 5.295e$-$02 && 72 + 23\sterm 2 D 5/\\ 
                 &   3/2   &    108263.47 &       43.32 & 5.17 & 5.266e$-$02 && 72 + 23\sterm 2 D 5/\\ 
  
\hline\\
\end{tabular}
\end{center}
\end{table*}


\clearpage

\begin{table}
\caption{Comparison of $J=5/2$ level separation, in cm$^{-1}$}
\label{separation}
\begin{center}
\begin{tabular}{r r r r r r}
\hline\hline
Term & Level & \quad & \multicolumn{3}{c}{Separation}\\
\cline{4-6}
     &       &\qquad &   Obs. & Garstang & Present \\
\hline
\Term 6 S {}/ & 0 \\
\Term 4 G {}/ & 32301 && 32301 & 32213 & 32260 \\
\Term 4 P {}/ & 35254 && 2953 & 2882 & 2969\\
\Term 4 D {}/ & 38935 && 3681 & 3372 & 3739 \\
\sterm 2 D 5/ & 49542 && 10606 & 10636 & 10615\\
\sterm 2 F 3/ & 52167 && 2625 & 2760 & 2649 \\
\Term 4 F {}/ & 52838 && 671 & 688 & 661 \\
\sterm 2 F 5/ & 61157 && 8319 & 7764 & 8235 \\
\sterm 2 D 3/ & 74133 && 12977 & 12919 & 13018 \\
\sterm 2 D 1/ & 108242 && 34109 & 33744 & 34073 \\
\hline
\end{tabular}
\end{center}
\end{table}

\clearpage


\def\qP{$^4\!P$}
\def\qD{$^4\!D$}
\def\qG{$^4\!G$}
\def\sS{$^6\!S$}

\begin{table*}
$^{~~~}$

\vskip-0.75truein

\caption{Comparison of normalized E2 transition rates $A_{J'J}$ (in
s$^{-1}$) in
emission for different theories. 1) from
Garstang (1958)
and 2) from Raassen \& Uylings (1997),
MCHF+BP and
MCDHF+B
Froese Fischer \& Rubin (1998)
and Present
(* denotes the previously recommended value).
}
\label{compare-E2}
\begin{center}
\small
\begin{tabular}{r r r r r l l l l l}
\hline\hline
$LS$ & $L'S'$ & $J$ & $J'$ & Obs. $E$&
\multicolumn{2}{c}{Semi-empirical}&   {\small MCHF+BP} & \small{MCDHF+B}
& Present\\
 & & & &(cm$^{-1}$) & \multicolumn{1}{c}{1)} & \multicolumn{1}{c}{2)}& \\
\hline
\qG & \qP & 7/2 & 5/2 & 2948.0 & 2.3(-09)& 3.42(-09)* & 5.90(-13)& 1.01(-18) & 3.68(-09)\\
    &     & 5/2 & 5/2 & 2952.8 & 1.4(-09)& 1.83(-09)* & 1.01(-13)& 6.99(-08) & 1.86(-09)\\
    &     & 9/2 & 5/2 & 2961.0 & 6.4(-10)& 1.75(-10)* & 2.63(-11)& 5.83(-08) & 7.50(-10)  \\
    &     & 7/2 & 3/2 & 3027.0 & 1.3(-09)& 6.22(-10)* & 4.61(-11)& 1.07(-08) & 1.60(-09)\\
    &     & 5/2 & 3/2 & 3031.8 & 8.5(-10)& 1.34(-09)* & 6.89(-12)& 7.38(-08) & 1.30(-09)\\
    &     &     & 1/2 & 3105.8 & 3.4(-10)& 9.79(-11)* & 2.91(-11)& 1.21(-06) & 4.02(-10)\\
\qP & \qD & 3/2 & 7/2 & 3446.0 & 2.9(-06)& 4.65(-06)  & 4.73(-06)*& 3.06(-09)& 4.97(-06)\\
    &     & 5/2 & 7/2 & 3525.0 & 4.9(-06)& 7.65(-06) & 8.02(-06)*& 3.16(-09) & 8.41(-06)\\
    &     & 1/2 & 5/2 & 3528.0 & 3.1(-06)& 5.08(-06) & 4.98(-06)*& 1.26(-07) & 5.21(-06)\\
    &     &     & 3/2 & 3531.0 & 1.8(-06)& 3.04(-06) & 2.98(-06)*& 3.19(-07) & 3.01(-06)\\
    &     & 3/2 & 1/2 & 3564.0 & 6.9(-06)& 1.23(-05) & 1.15(-05)*& 9.70(-08) & 1.17(-05)\\
    &     &     & 5/2 & 3602.0 & 1.5(-07)& 2.26(-07) & 2.13(-07)*& 5.22(-07) & 2.46(-07)\\
    &     &     & 3/2 & 3605.0 & 3.3(-06)& 5.45(-06) & 5.17(-06)*& 2.04(-06) & 5.52(-06)\\
    &     & 5/2 & 1/2 & 3643.0 & 9.9(-07)& 1.65(-06) & 1.68(-06)*& 1.53(-07) & 1.64(-06)\\
     &     &     & 5/2 & 3681.0 & 5.5(-06)& 8.64(-06) & 8.59(-06)*& 7.11(-06) & 9.17(-06)\\
    &     &     & 3/2 & 3684.0 & 3.4(-06)& 5.65(-06) & 5.48(-06)*& 4.20(-07)  & 5.60(-06)\\
\qG & \qD & 7/2 & 7/2 & 6473.0 & 6.8(-07)& 1.17(-06) & 3.62(-10)*& 4.97(-07)  & 1.11(-06)\\
    &     & 5/2 & 7/2 & 6477.8 & 5.2(-08)& 8.29(-08) & 6.45(-11)*& 3.48(-08) & 7.42(-08)\\
    &     & 9/2 & 7/2 & 6486.0 & 1.5(-06)& 2.85(-06) & 2.61(-09)*& 1.49(-06) & 2.77(-06)\\
    &     & 11/2 & 7/2& 6533.5 & 1.7(-07)& 2.96(-10) & 1.75(-08)*& 1.20(-06) & 2.59(-07)\\
    &     & 5/2 & 1/2 & 6595.8 & 1.0(-07)& 2.84(-08) & 1.67(-08)*& 1.56(-06) & 1.21(-07)\\
    &     & 7/2 & 5/2 & 6629.0 & 1.1(-06)& 2.28(-06) & 4.31(-09)*& 4.91(-07) & 1.99(-06)\\
    &     &     & 3/2 & 6632.0 & 6.0(-07)& 2.08(-07) & 1.26(-08)*& 1.03(-07) & 9.19(-07)\\
    &     & 5/2 & 5/2 & 6633.8 & 7.2(-07)& 1.28(-06) & 4.84(-10)*& 1.07(-07) & 1.20(-06)\\
    &     &     & 3/2 & 6636.8 & 6.6(-07)& 1.52(-06) & 4.95(-09)*& 2.21(-07) & 1.16(-06)\\
    &     & 9/2 & 5/2 & 6642.0 & 6.7(-07)& 2.76(-07) & 1.43(-08)*& 8.83(-07) & 1.09(-06)\\
\sS & \qG & 5/2 & 9/2 & 32293.0 & $<$1.0(-09)& 6.92(-12)*& 1.32(-09)& 1.27(-05) & 1.85(-09)\\
    &     &     & 5/2 &  32301.2 & $<$1.0(-09)& 1.95(-08)*& 4.07(-11)& 6.09(-05) & 2.30(-08)\\
&     &     & 7/2 &  32306.0 & $<$1.0(-09)& 3.18(-08)*&5.72(-12)& 1.10(-05) & 3.32(-08)\\
    & \qP & 5/2 & 5/2 &  35254.0 & 3.9(-05)& 4.26(-05) & 3.10(-05)*& 4.01(-03) & 4.02(-05)\\
    &     &     & 3/2 &  35333.0 & 1.5(-05)& 1.99(-05) & 8.84(-06)*& 5.83(-01) & 1.21(-05)\\
    &     &     & 1/2 &  35407.0 & v.s.    & 1.67(-06) & 2.10(-07)*& 1.28(~00) & 1.47(-07)\\
    & \qD & 5/2 & 7/2 &  38779.0 & 1.1(-03)& 1.13(-03) & 1.18(-03)*& 4.40(-04) & 1.27(-03)\\
    &     &     & 1/2 &  38897.0 & 1.8(-04)& 1.97(-04) & 2.03(-04)*& 1.58(-02) & 2.17(-04)\\
    &     &     & 5/2 &  38935.0 & 1.0(-03)& 1.09(-03) & 1.15(-03)*& 1.58(-03) & 1.23(-03)\\
    &     &     & 3/2 &  38938.0 & 6.2(-04)& 6.72(-04) & 6.86(-04)*& 2.64(-02) & 7.33(-04)\\
\hline
\end{tabular}
\end{center}
\noindent
Note.-- v.s. means very small in Garstang (1958) compilation.

\end{table*}

\clearpage


\begin{table*}
\caption{Comparison of normalized M1 transition rates $A_{J'J}$ (in
s$^{-1}$) in
emission for different theories. 1) from
Garstang (1958) and 2) from Raassen \& Uylings (1997),
MCHF+BP and
MCDHF+B
Froese Fischer \& Rubin (1998)
and Present
(* denotes the previously recommended value).
}
 \label{compare-M1}
 \begin{center}
 \small
 \begin{tabular}{r r r r r l l l l l}
 \hline\hline
 $LS$ & $L'S'$ & $J$ & $J'$ & Obs. $E$&
 \multicolumn{2}{c}{Semi-empirical}&   {\small MCHF+BP} & \small{MCDHF+B}
 & Present\\
 & & & &(cm$^{-1}$) & \multicolumn{1}{c}{1)} & \multicolumn{1}{c}{2)}& \\
  \hline
\qG & \qP & 7/2 & 5/2 &  2948.0 & 1.5(-05)& 1.89(-05) & 8.02(-08)& 1.51(-05)* & 2.37(-05)\\
      &     & 5/2 & 5/2 &  2952.8 & 6.8(-05)& 1.04(-04) & 7.36(-07)& 7.90(-05)* & 1.32(-04)\\
      &     & 5/2 & 3/2 & 3031.8 & 8.6(-06)& 1.33(-05) & 1.21(-07)& 9.79(-06)* & 1.71(-05)\\
\qP & \qD & 1/2 & 1/2 & 3490.0 & 5.8(-02)& 6.51(-02) & 6.04(-02)*& 4.94(-02) & 7.07(-02)\\
     &     & 5/2 & 7/2 & 3525.0 & 3.8(-02)& 4.26(-02) & 3.86(-02)*& 3.31(-02) & 4.56(-02)\\
     &     & 1/2 & 3/2 & 3531.0 & 1.4(-03)& 1.89(-03) & 1.70(-03)*& 1.48(-03) & 2.20(-03)\\
     &     & 3/2 & 1/2 & 3564.0 & 3.4(-02)& 4.00(-02) & 3.67(-02)*& 3.08(-02) & 4.47(-02)\\
     &     &     & 5/2 &  3602.0 & 3.5(-03)& 3.74(-03) & 3.59(-03)*& 2.92(-03) & 4.47(-03)\\
     &     &     & 3/2 &  3605.0 & 3.9(-02)& 4.30(-02) & 3.97(-02)*& 3.32(-02) & 4.73(-02)\\
     &     & 5/2 & 5/2 & 3681.0 & 2.2(-02)& 2.46(-02) & 2.26(-02)*& 1.92(-02) & 2.62(-02)\\
     &     &     & 3/2 & 3684.0 & 1.8(-02)& 2.06(-02) & 1.83(-02)*& 1.61(-02) & 2.11(-02)\\
\qG & \qD & 7/2 & 7/2 &  6473.0 & 7.6(-04)& 1.10(-03) & 2.12(-05)& 8.29(-04)* & 1.33(-03)\\
      &     & 5/2 & 7/2 & 6477.8 & 7.6(-05)& 1.08(-04) & 2.18(-06)& 8.18(-05)* & 1.33(-04)\\
      &     & 9/2 & 7/2 & 6486.0 & 6.3(-04)& 9.02(-04) & 1.46(-05)& 6.89(-04)* & 1.10(-03)\\
      &     & 7/2 & 5/2 &  6629.0 & v.s.    & 2.51(-08) & 6.29(-09)& 2.66(-04)* & 1.68(-08)\\
    &     & 5/2 & 5/2 &  6633.8 & 5.8(-04)& 8.64(-04) & 1.90(-05)& 6.82(-04)* & 1.01(-03)\\
   &     &     & 3/2 &  6636.8 & 2.8(-04)& 3.69(-04) & 7.92(-06)& 2.82(-04)* & 4.37(-04)\\
\sS & \qG & 5/2 & 5/2 &   32301.2 & 1.0(-05)& 1.53(-05) & 8.78(-08)& 1.01(-05)* & 2.16(-05)\\
   &     &     & 7/2 & 32306.0 & $<$1.0(-07)& 3.78(-08) & 5.28(-10)& 1.42(-08)* & 5.72(-08)\\
   & \qP & 5/2 & 5/2 & 35254.0 & 1.4& 1.42 & 1.53*& 1.17 & 1.56\\
&     &     & 3/2 &  35333.0 & 8.8(-01)& 9.23(-01) & 9.98(-01)*& 7.59(-01) & 1.02\\
& \qD & 5/2 & 7/2 &38779.0 & 2.0(-04)& 5.90(-04) & 7.18(-04)*& 2.44(-04) & 7.67(-04)\\
    &      &     & 5/2 & 38935.0 & 5.1(-02)& 5.69(-02) & 5.32(-02)*& 3.40(-02) & 6.64(-02)\\
    &     &     & 3/2 &  38938.0 & 3.8(-02)& 2.75(-02) & 2.61(-02)*& 1.65(-02) & 3.33(-02)\\
\hline
\end{tabular}
\end{center}
\noindent
Note.-- v.s. means very small in Garstang (1958) compilation.

\end{table*}

\clearpage

\begin{deluxetable}{lrrllllll}
\tabletypesize{\scriptsize}
\tablecaption{Comparison with the Garstang values of some $A_{ki}$ 
for E2 and M1 transitions between higher levels.}

\tablewidth{0pt}
\tablehead{
\colhead{Multiplet} & $g_i$ & $g_k$ &\quad\quad & 
\multicolumn{2}{c}{E2}  & & \multicolumn{2}{c}{M1}\\
\cline{5-6} \cline{8-9}
	  &       &       &&  Present & Garstang    & & Present & Garstang}
\def\Term#1 #2 #3/{\mbox{$\,^{#1}\!#2_{#3}$ }}
\def\Termo#1 #2 #3/{\mbox{$\,^{#1}\!#2^o_{#3}$ }}
\def\sterm #1 #2 #3/{\mbox{$\,_{#3}\!^{#1}\!#2$}}
\columnsep=.5in
\startdata
\Term 4 G {}/ -- \Term 2 I {}/  & 8 & 12 && 1.14(-06) & 1.0(-06) & \\
				& 10 & 12 && 2.61(-06) & 3.3(-07) & & 2.32(-03) & 3.0(-03) \\
				& 10 & 14 && 8.81(-07) & 1.6(-06) & \\
				& 12 & 12 && 3.28(-07) & 3.2(-07) & & 7.54(-03) & 5.1(-03) \\
				& 12 & 14 && 7.65(-06) & 7.1(-06) & & 2.85(-04) & 1.6(-04) \\

\Term 4 G {}/ -- $_5\!{\!^2\!D}$ & 6 & 4 && 5.96(-03) & 3.7(-03) & & 0.0151 & 0.010 \\
				 & 6 & 6 && 3.70(-04) & 3.3(-04) & & 0.176 & 0.13 \\
				 & 8 & 4 && 2.09(-03) & 1.3(-03) & \\
				 & 8 & 6 && 7.37(-04) & 5.9(-04) & & 0.171 & 0.12 \\
				 & 10 & 6 && 2.65(-04) & 2.1(-04) & \\

\Term 4 G {}/ -- $_3\!{\!^2\!F}$ & 6 & 6 && 0.0134 & 4.5(-03) & & 0.140 & 0.21 \\
				& 6 & 8 && 9.04(-05) & 2.7(-05) & & 0.0155 & 0.015 \\
				& 8 & 6 && 0.0229 & 8.1(-03) & & 0.463 & 0.47 \\
				& 8 & 8 && 8.43(-04) & 2.3(-04) & & 0.0825 & 0.080 \\
				& 10 & 6 && 0.0120 & 4.1(-03) & \\
				& 10 & 8 && 1.57(-03) & 4.6(-04) & & 0.569 & 0.50 \\
				& 12 & 8 && 5.21(-04) & 1.4(-04) & \\

\Term 4 G {}/ -- \Term 4 F {}/  & 6 & 4 && 0.195 & 0.18 & &0.157 & 0.14 \\
				& 8 & 4 && 0.0788 & 0.071 & \\
				& 6 & 6 && 0.0621 & 0.064 && 0.287 & 0.19 \\
				& 8 & 6 && 0.109 & 0.11 && 0.0209 & 9.4(-04) \\
				& 10 & 6 && 0.0594 & 0.061 \\
				& 6 & 8 && 6.59(-03) & 4.1(-03)& & 0.0156 & 0.014 \\
				& 8 & 8 && 0.0713 & 0.067 && 0.156 & 0.13 \\
				& 10 & 8 && 0.155 & 0.15 && 0.0514 & 0.039 \\
				& 12 & 8 && 0.0436 & 0.040 \\
				& 6 & 10 && 8.88(-05) & 8.7(-05) \\
				& 8 & 10 && 3.92(-03) & 3.8(-03)& & 0.0125 & 0.0023 \\
				& 10 & 10 && 0.0474 & 0.045 && 0.0733 & 0.073 \\
				& 12 & 10 && 0.223 & 0.21 && 0.160 & 0.11 \\

\Term 4 G {}/ -- \Term 2 H {}/  & 6 & 10 && 5.32(-08) & 1.9(-06) & \\
				& 8 & 10 && 4.40(-05) & 3.4(-05) & & 0.179 & 0.13 \\
				& 8 & 12 && 8.41(-07) & {\small v.s.} & \\
				& 10 & 10 && 5.52(-04) & 4.8(-04) & & 0.201 & 0.042 \\
				& 10 & 12 && 1.07(-07) & 7.8(-05) & & 0.228 & 0.61 \\
				& 12 & 10 && 2.12(-03) & 1.3(-03) & & 0.0150 & 0.56 \\
				& 12 & 12 && 6.30(-07) & {\small v.s.} & & 0.554 & 0.47 \\

\Term 4 P {}/ -- $_5\!{\!^2\!D}$ & 2 & 4 && 2.50(-05) & 7.0(-06) & & 0.0718 & 0.065 \\
				 & 2 & 6 && 2.24(-07) & 1.1(-07) & \\
				 & 4 & 4 && 4.89(-05) & 2.0(-05) & & 0.280 & 0.25 \\
				 & 4 & 6 && 8.94(-06) & 5.9(-06) & & 0.112 & 0.10 \\
				 & 6 & 4 && 1.82(-05) & 7.3(-06) & & 0.0763 & 0.069 \\
				 & 6 & 6 && 8.50(-06) & 2.8(-06) & & 0.594 & 0.54 \\

\Term 4 P {}/ -- $_3\!{\!^2\!F}$ & 2 & 6 && 1.02(-05) & 2.4(-05) & \\
				 & 4 & 6 && 1.32(-04) & 1.5(-05) & & 0.0157 & 0.013 \\
				 & 4 & 8 && 2.42(-05) & 2.0(-05) & \\
				 & 6 & 6 && 2.79(-04) & 7.1(-05) & & 0.101 & 0.087 \\
				 & 6 & 8 && 7.97(-06) & 2.5(-07) & & 5.87(-04) & 5(-04) \\

\Term 4 D {}/ -- $_3\!{\!^2\!F}$ & 2 & 6 && 1.28(-03) & 5.4(-04) & \\
				 & 4 & 6 && 7.19(-04) & 3.0(-04) & & 0.0128 & 0.021 \\
				 & 4 & 8 && 6.91(-05) & 2.3(-05) & \\
				 & 6 & 6 && 1.91(-03) & 7.1(-04) & & 4.85(-03) & 0.022 \\
				 & 6 & 8 && 8.34(-05) & 2.8(-05) & & 0.0405 & 0.034 \\
				 & 8 & 6 && 4.62(-04) & 1.9(-04) & & 8.72(-05) & 7.2(-04) \\
				 & 8 & 8 && 9.54(-05) & 3.3(-05) & & 0.0994 & 0.10 \\ 

\Term 4 D {}/ -- \Term 4 F {}/  & 2 & 4 && 7.62(-03) & 8.3(-03) & & 0.107 & 0.10 \\
				& 2 & 6 && 6.67(-03) & 7.7(-03) & \\
				& 4 & 4 && 0.0151 & 0.016 & & 0.177 & 0.16 \\
				& 4 & 6 && 3.35(-03) & 4.1(-03) & & 0.0327 & 0.023 \\
				& 4 & 8 && 8.26(-03) & 8.5(-03) & \\
				& 6 & 4 && 4.34(-03) & 4.5(-03) & & 0.0432 & 0.039 \\
				& 6 & 6 && 0.0110 & 0.012 & & 0.216 & 0.18 \\
				& 6 & 8 && 8.82(-03) & 9.2(-03) & & 8.32(-03) & 9.4(-03) \\
				& 6 & 10 && 5.79(-03) & 5.7(-03) & \\
				& 8 & 4 && 1.92(-04) & 1.9(-04) & \\
				& 8 & 6 && 1.95(-03) & 2.1(-03) & & 0.0330 & 0.028 \\
				& 8 & 8 && 0.0102 & 0.011 & & 0.120 & 0.10 \\
				& 8 & 10 && 0.0221 & 0.023 & & 0.151 & 0.14 \\

\Term 4 F {}/ -- $_5\!{\!^2\!F}$ & 4 & 6 && 5.08(-08) & 6.8(-08) & & 0.111 & 0.098 \\
				 & 6 & 6 && 1.34(-04) & 3.1(-05) & & 0.0200 & 0.016 \\
				 & 8 & 6 && 3.54(-06) & 7.2(-07) & & 0.0399 & 0.027 \\
				 & 10 & 6 && 2.18(-06) & 1.1(-07) \\
				 & 4 & 8 && 1.52(-08) & 3.3(-09) \\
				 & 6 & 8 && 2.02(-05) & 5.4(-06) & & 0.0242 & 0.026 \\
				 & 8 & 8 && 2.78(-05) & 7.0(-06) & & 0.0333 & 0.013 \\
				 & 10 & 8 && 7.12(-08) & 4.7(-08) & & 0.129 & 0.11\\

$_5\!{\!^2\!G}$ -- $_3\!{\!^2\!G}$ & 8 & 8 && 0.460 & 0.40 & & 2.56(-04) & {\small v.s.}\\
				   & 10 & 8 && 0.0335 & 0.028 & & 0.277 & 0.0053 \\
				   & 8 & 10 && 0.0371 & 0.033 & & 0.258 & 0.029 \\
				   & 10 & 10 && 0.365 & 0.29 & & 1.71(-04) & 0.017\\

$_5\!{\!^2\!F}$ -- $_3\!{\!^2\!D}$ & 6 & 4&& 7.29(-03) & 7.1(-03) & & 0.102 & 0.10 \\
				   & 8 & 4 && 1.38(-03) & 1.3(-03) \\
				   & 6 & 6 && 1.11(-03) & 1.3(-03) & & 0.148 & 0.18 \\
				   & 8 & 6 && 6.87(-03) & 7.2(-03) & & 0.103 & 0.10 \\

$_5\!{\!^2\!F}$ -- $_3\!{\!^2\!G}$ & 6 & 8 && 0.0570 & 0.055 & & 0.107 & 0.091 \\
				   & 6 & 10 && 3.78(-03) & 2.9(-03) & \\
				   & 8 & 8 && 8.66(-03) & 7.1(-03) & & 0.284 & 0.22 \\
				   & 8 & 10 && 0.0583 & 0.058 & & 0.109 & 0.10 \\

\enddata

\noindent
Note.-- v.s. means very small in Garstang (1958) compilation.

\end{deluxetable}

\clearpage

\begin{figure}
\vskip-1.truein
\plotfiddle{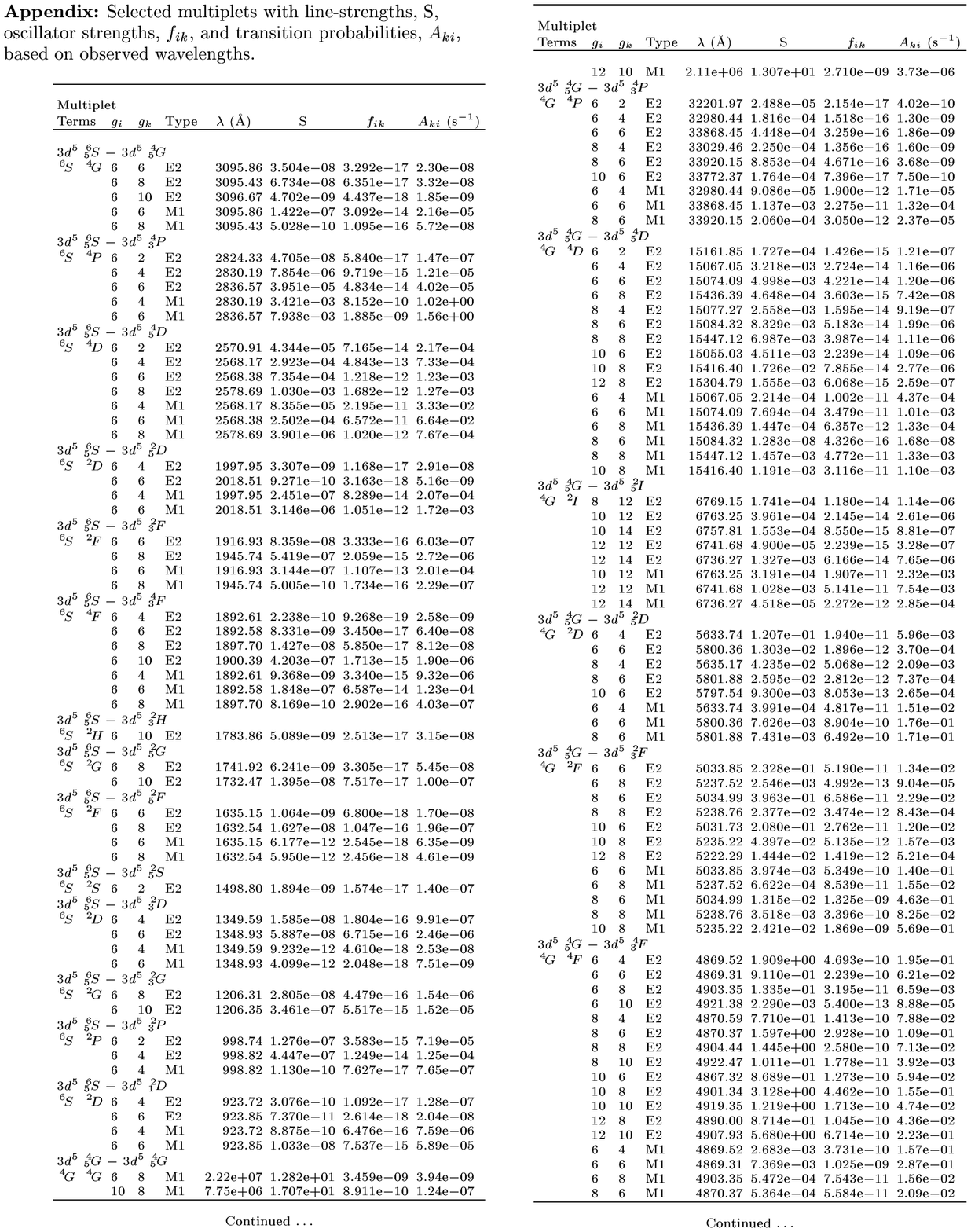}{0.1truein}{0}{570}{713}{-40}{-450}
\vskip-0.1truein
\end{figure}

\clearpage

\begin{figure}
\vskip-1.truein
\plotfiddle{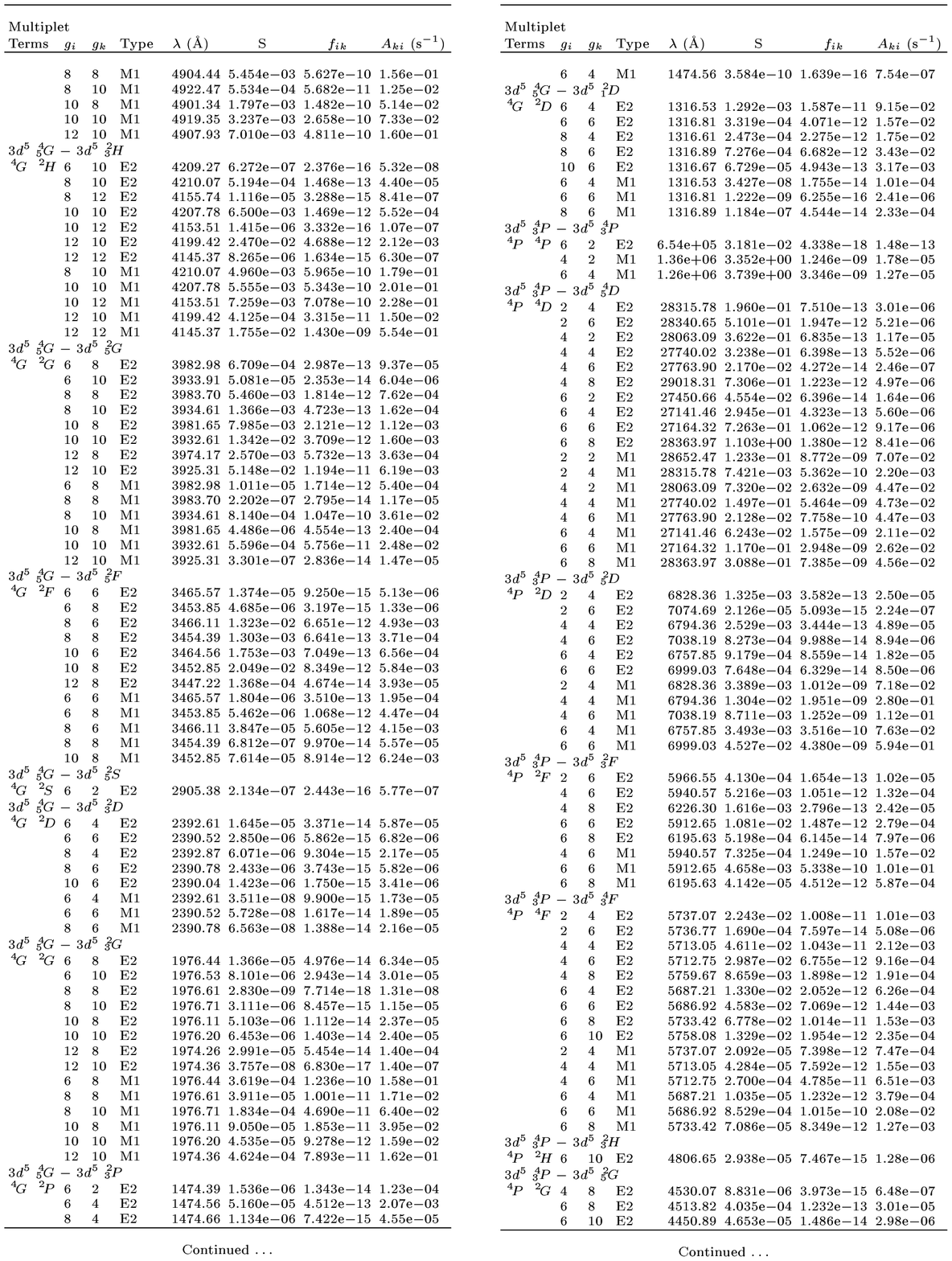}{0.1truein}{0}{570}{713}{-40}{-450}
\vskip-0.1truein
\end{figure}

\clearpage

\begin{figure}
\vskip-1.truein
\plotfiddle{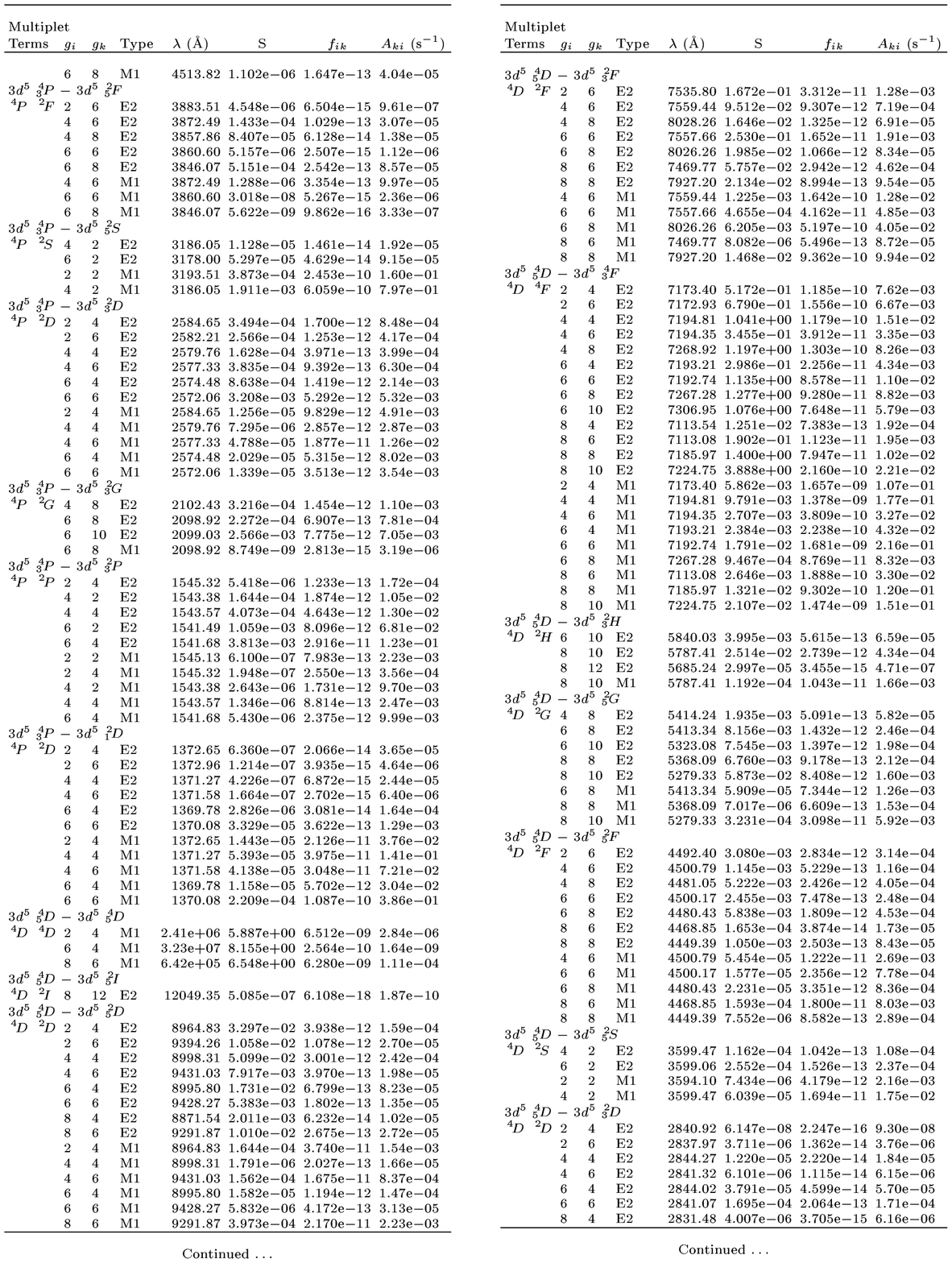}{0.1truein}{0}{570}{713}{-40}{-450}
\vskip-0.1truein
\end{figure}

\clearpage

\begin{figure}
\vskip-1.truein
\plotfiddle{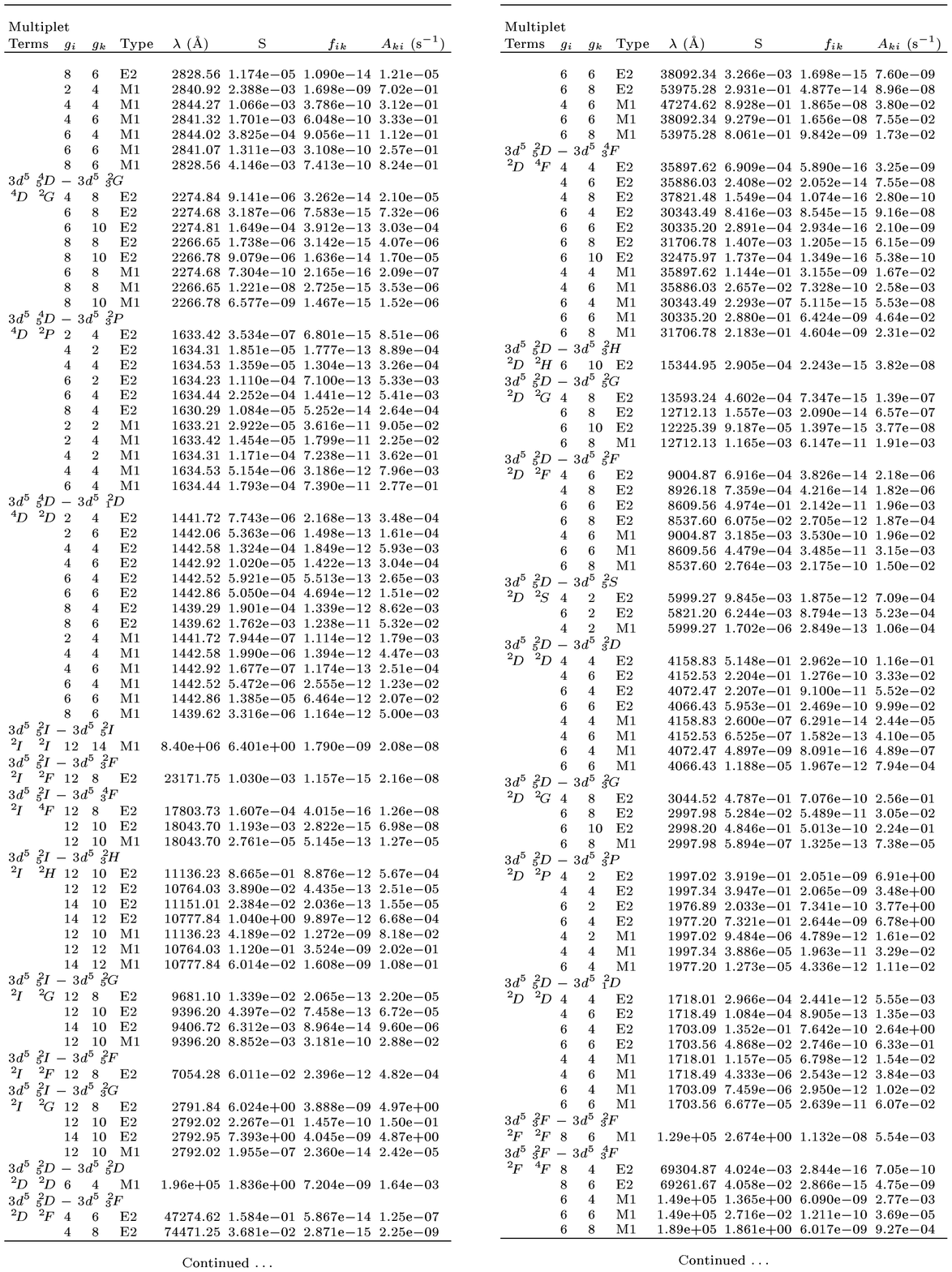}{0.1truein}{0}{570}{713}{-40}{-450}
\vskip-0.1truein
\end{figure}

\clearpage

\begin{figure}
\vskip-1.truein
\plotfiddle{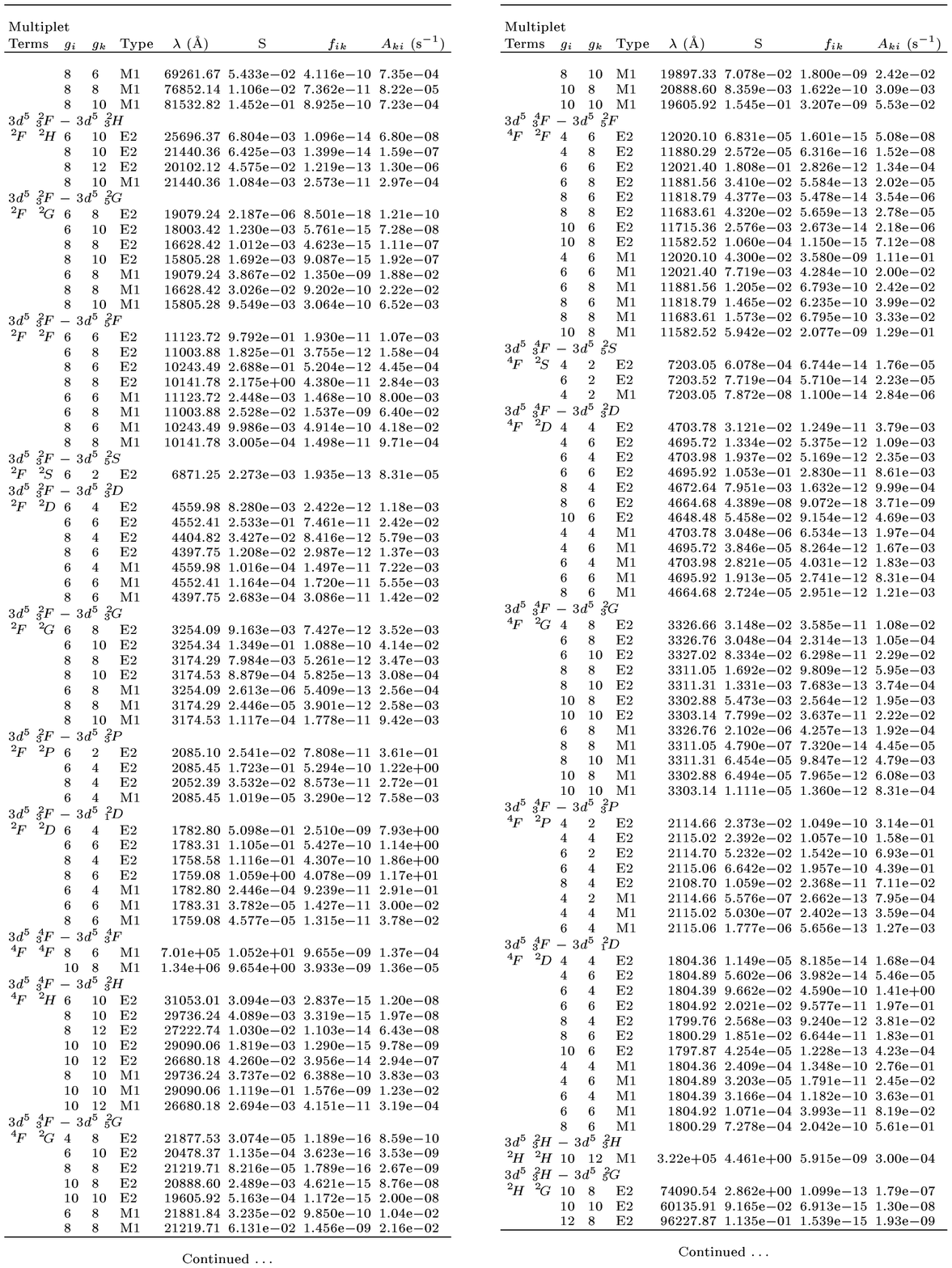}{0.1truein}{0}{570}{713}{-40}{-450}
\vskip-0.1truein
\end{figure}

\clearpage

\begin{figure}
\vskip-1.truein
\plotfiddle{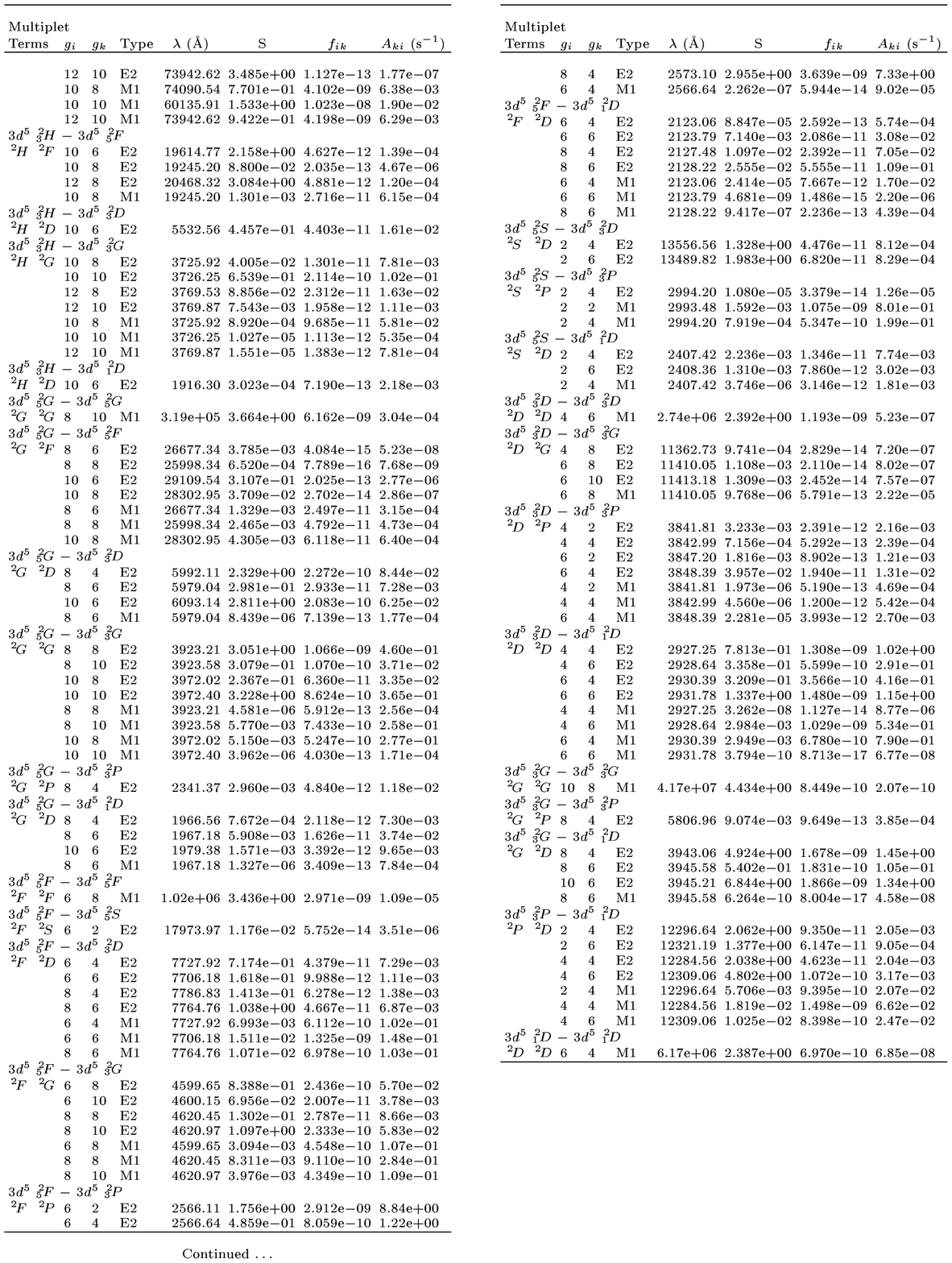}{0.1truein}{0}{570}{713}{-40}{-450}
\vskip-0.1truein
\end{figure}

\end{document}